# Quasar Viscosity Crisis

Andy Lawrence, Institute for Astronomy, SUPA (Scottish Universities Physics Alliance), Royal Observatory, Blackford Hill, Edinburgh EH9 3HJ, United Kingdom (email: al@roe.ac.uk)

*Recent observations of extreme variability in Active Galactic Nuclei have pushed standard viscous accretion disc models over an edge. "Extreme reprocessing" where an erratically variable central quasi-point source is entirely responsible for heating an otherwise cold and passive low-viscosity disc, may be the best route forward.*

It is widely believed that Active Galactic Nuclei (AGN), including the most luminous examples, the quasars, are powered by accretion discs surrounding supermassive black holes. We have understood the general principles of accretion discs since the 1970s (refs 1,2,3). The disc is differentially rotating, so neighbouring rings are slipping past each other. Some viscous process causes a drag between the rings, transferring angular momentum outwards, and producing local heating. If that local heating is also radiated thermally on the spot, this process determines the radial temperature profile ($T \propto R^{-3/4}$). A further simplifying assumption, that viscosity is proportional to sound speed, allows a complete solution of the disc structure. A well known problem is that standard molecular viscosity, where particles from the fast-lane slip into the slow lane and vice versa, is far too weak a process, producing a very small luminosity. From the 1970s onwards it was widely assumed that some of kind of turbulence and/or magnetic stresses would produce a viscosity-like effect. This idea was put on a sound footing in 1991, with the development of the theory of magneto-rotational instability (MRI – ref 4).

Accretion discs nicely explain the luminosity and compactness of AGN, and also the observed peak of the spectral energy distribution (SED) in the UV. There have always been difficulties getting the details right (e.g ref 5), but these problems may be explained by effects which modify the SED, such as a Comptonising atmosphere, or a system of clouds surrounding the disc (refs 5,6). However by far the worst problem is variability. AGN vary significantly on timescales of weeks to months, whereas discs with the right degree of viscosity to explain the luminosity should take thousands of years to change their optical emission. Furthermore, variations at different wavelengths, from the optical through to the UV, vary simultaneously, with the peaks lined up (ref 7; see Fig. 1a below), whereas in an accretion disc, different wavelengths come from different radii; changes should propagate through the disc.

This situation was rescued in the 1990s by the idea of X-ray reprocessing (e.g. ref 8). The central X-ray source, which can vary much more quickly than the part of the disc making optical light, shines on the disc and heats it. At any radius, heating has two causes - viscous heating, which changes only slowly, and X-ray heating, which can change quickly. Noticeably, although the shortest UV wavelengths might change by say a factor two peak-to-trough, the redder optical wavelengths change only by a few percent. There have been many papers arguing about whether or not X-ray reprocessing works in detail. The strongest argument in favour is the observation of delays between the variations at different wavelengths - on a timescale of hours-to-days, consistent with light travel time delays (eg refs 9,10).

However, the variability problem is now reaching a new crisis, with the observation of *extreme variability* in some objects - factors of several over a decade or so, including, crucially, at optical wavelengths, not just in the extreme UV or in X-rays. Large changes have been known in a handful of nearby low-luminosity AGN for many years, but comparison of the data from the Sloan Digital Sky Survey (SDSS) and the Panoramic Survey Telescope and Rapid Response System (PanSTARRS) has revealed large numbers of such objects (e.g. ref 11; see Fig. 1b below), including many at high luminosity. These objects have generally been referred to as "changing look quasars". The broad emission lines (BELs) that normally accompany Type I, (i.e. quasar-like) AGN seem to come and go along with the optical continuum; when the continuum and broad lines plummet, what is left behind is the narrow emission lines that dominate Type II AGN. The changing BELs tell us that the unseen far-UV must be dramatically changing, as well as the observed optical emission.

Because these large changes occur in optical emission, not just in X-ray or far UV emission, it seems hard to avoid the conclusion that the outer parts of the disc itself is undergoing a gross physical change on a timescale inconsistent with viscous heating. Furthermore, recent work, such as that comparing DES and SDSS surveys, seems to suggest that extreme variability is not that unusual – possibly 30-50% of quasars vary by a large amount sometimes (ref 12). Also, studies of the variability structure function seem to suggest that the degree of optical variability for a typical quasar climbs inexorably to longer timescales (ref 13). Although some AGN have larger typical variability than others on any given timescale, it seems likely that all AGN vary dramatically if you just wait long enough.

One might wonder whether some kind of variable obscuration – e.g. passing clouds in the clumpy torus – can explain the variability. However, papers studying large changes usually conclude that this idea doesn't fit the observations – the timescales, the (lack of) colour changes, and the relative line and continuum changes look wrong (see Fig. 1b). It seems that we really have to confront the fact that accretion disc models are failing. Of course, good theorists have long known that standard viscous accretion disc theory is just too simple – it's just that it remains the observers' paradigm. When interpreting data, authors routinely assume that standard theory is correct, and write optimistically of "accretion disc instabilities" to explain outbursts. The problem is that the existence of common large amplitude variability suggests that discs are in a state of permanent exception; it is not reasonable to describe them with standard viscous accretion discs at all. As Pringle said way back in 1981 (ref 14), in accretion disc physics, "instability" really means "inconsistency".

We cannot solve this problem by simply cranking up the viscosity parameter. The rate of torque is closely related to the viscous scale length and so the disc height, as well as the assumption of local transport and heating; the disc approximation breaks down completely. There are two likely routes forward:

**Route–1: Extreme reprocessing**. The simplest picture is that a disc is present, but has very low viscosity, and is entirely passive - it is cold unless externally heated. All the gravitational energy emerges from a central quasi-point source - in the extreme UV and/or X-rays - which heats the passive disc. This central source could be an inner region in quasi-spherical turbulent accretion; a region where the spin-energy of the black hole is extracted by the Blandford-Znajek process; or an inner region at around say 3-10 Schwarzschild radii (3-10 $R_s$) where there really is a viscous disc, with much shorter timescale than in the outer disc at say 30-100 $R_s$.

This picture is appealing because of the light-travel time delays seen in some objects. However, the outer disc may be very dense and massive, and before it is heated by any central source, it is cold. The response of the disc to erratic variations of the central source will therefore be complicated, having both a prompt response (skin heating and scattering) and a smoothed response (deep heating from the history of the central source luminosity). Also, the well known problems remain with the peak of the SED, the optical source size, and the relative amplitude of variability at different wavelengths. It is possible that these problems can be solved by a surrounding cloud system which does most of the reprocessing. It could be that the main role of the disc is to generate reprocessing clumps – all the way from the inner disc at 10 $R_s$ out to the traditional Broad Line Region at 1000 $R_s$.

**Route–2: non-local processes**. On the other hand, perhaps we have to abandon the hope that the transfer of angular momentum, the generation of heat, and the radiation of that heat, can all be approximated as local and co-located processes. Large-scale magnetic fields can cause one ring to drag on a very distant ring; corkscrew-like outflows can carry angular momentum away; if infall on a dynamical timescale is possible, heating and radiation may be only loosely coupled to gravitational energy generation. All these ideas have the smell of physical realism. Much of the basic physics was laid out in the classic review by Rees (ref 15). The trouble is that there are too many such ideas, and most of them sound horribly difficult to work out in detail - what would we expect the SED to look like?

On the principle of looking for our lost keys under the streetlamp, then, although we are forced to abandon standard viscous disc theory, our best bet is perhaps to work as hard as we can on Route–1 until we see if it fails us too.

## **References**

(1) Pringle, J.E., and Rees, M.J., 1972. A&A, 21, 1.
(2) Shakura, N.I., and Sunyaev, R.A., 1973. A&A, 24, 337.
(3) Frank, J., King, A., and Raine, D., Accretion Power in Astrophysics, Cambridge University Press.
(4) Balbus, S.A., and Hawley, J.H., 1991. ApJ, 376, 214.
(5) Lawrence, A., 2012. MNRAS, 423, 451.
(6) Gardner, E., and Done, C., 2017 MNRAS, 470, 3591.
(7) Clavel, J., et al 1991. ApJ, 366, 94.
(8) Clavel, J., et al 1992. ApJ, 393, 113.
(9) Edelson, R., et al, 2015. ApJ 806,129
(10) McHardy, I., et al 2016. AN, 337, 4.
(11) Macleod, C.L., et al 2016. MNRAS, 457, 389.
(12) Rumbaugh, N. et al, 2017. ApJ in press (ArXiV 1706.0875)
(13) Morganson, E., 2014. ApJ 784, 92.
(14) Pringle, J.E.,1981. Ann.Rev.Astron.Astrophys., 19, 137.
(15) Rees, M.J, 1984. Ann.Rev.Astron.Astrophys., 22, 471

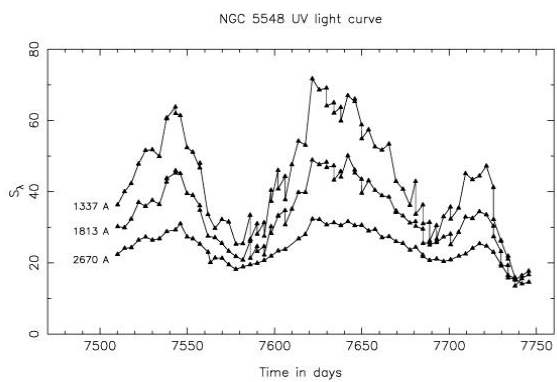

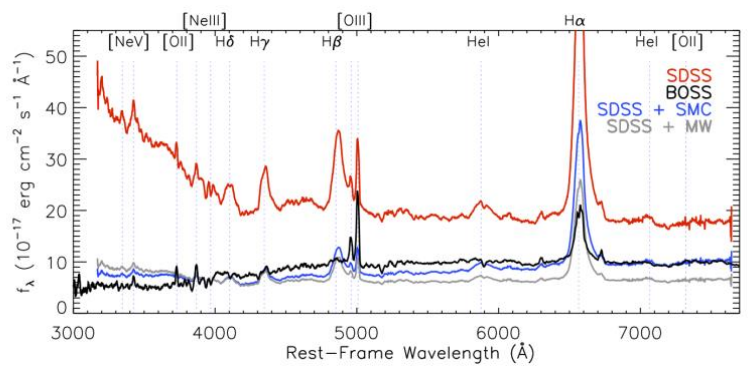

**Fig. 1a** *Variations in the near-UV brightness in NGC5548 at three different wavelengths, showing the short timescale, the simultaneity at different wavelengths, and the differing amplitude at different wavelengths – all serious problems for standard viscous disc theory. The data points were taken from the classic study of Clavel et al 1991*

**Fig. 1b** *A dramatic change over a period of years in the quasar J1021+1645, from MacLeod et al 2016. In the lower part of the plot, the black curve is the data. The blue and grey curves are (failed) attempts to model the collapse by a change in extinction.*